\documentclass[
    ,final            
  ]
  {aipproc}

\layoutstyle{6x9}
\newcommand{\etal}{\emph{et~al.}}
\newcommand{\aj}{\textit{Astron.~J.}}
\newcommand{\apj}{\textit{Astrophys.~J.}}
\newcommand{\apjl}{\textit{Astrophys.~J.~Lett.}}
\newcommand{\apjs}{\textit{Astrophys.~J.~Supp.~Ser.}}

\begin{document}

\title{Astronomical imaging: The theory of everything}


\author{David W. Hogg}{
  address={Center for Cosmology and Particle Physics,
           Department of Physics,
           New York University,\\
           4 Washington Place, New York, New York 10012, USA}
}
\author{Dustin Lang}{
  address={Department of Computer Science,
           University of Toronto,\\
           6 King's College Road, Toronto, Ontario, M5S~3G4 Canada}
}

\begin{abstract}
  We are developing automated systems to provide homogeneous
  calibration meta-data for heterogeneous imaging data, using the
  pixel content of the image alone where necessary.  Standardized and
  complete calibration meta-data permit generative modeling: A good
  model of the sky through wavelength and time---that is, a model of
  the positions, motions, spectra, and variability of all stellar
  sources, plus an intensity map of all cosmological sources---could
  synthesize or generate any astronomical image ever taken at any time
  with any equipment in any configuration.  We argue that the best-fit
  or highest likelihood model of the data is also the best possible
  astronomical catalog constructed from those data.  A generative
  model or catalog of this form is the best possible platform for
  automated discovery, because it is capable of identifying
  informative failures of the model in new data at the pixel level, or
  as statistical anomalies in the joint distribution of residuals from
  many images.  It is also, in some sense, an astronomer's ``theory of
  everything''.
\end{abstract}

\maketitle

\section{Catalogs and image models}

Astronomers love catalogs; we have been creating catalogs of celestial
sources for as long as there has been information recorded in hard
copy.  Why?  Consider two examples.

The first is Abell~\cite{abell}.  He spent thousands of hours
poring over images of the sky; his Catalog communicated information he
found in those images, so that other workers would not have to repeat
the effort. This was at a time when you couldn't just ``send them the
data and the code''.  Indeed, Abell's Catalog wasn't constructed using
code at all; there was no way to re-run the experiment, so the
experiment \emph{had} to be recorded and published in the form of the
output catalog.

The second is the Sloan Digital Sky Survey (SDSS)~\cite{sdss}.  Why
did the SDSS produce a catalog rather than just releasing one
enormous, five-band, terapixel image? The SDSS Collaboration produced
a catalog because investigators want to search for sources and measure
the fluxes of those sources, and they do this in a limited number of
ways.  The SDSS made it easier for them by pre-computing all these
fluxes and positions and making the computation output searchable.

Importantly, however, unlike Abell, the SDSS \emph{could have}
produced a piece of fast code and a fast interface to the data, and
\emph{could have} made it easy to run the code on the data and search
its output.  For any user, that would have been no worse; it could
even have been \emph{identical} in user interface and query language
(though---admittedly---much harder to implement at the present day).

Astronomers use the SDSS Catalogs (instead of working directly with
the image pixels) not just because they are easy to use but also
because they contain all of the SDSS Collaboration's \emph{knowledge}
about the data, encoded as proper data analysis procedures and error
estimates; this knowledge takes considerable time and effort to learn
and implement.  This is a very important aspect of a good catalog: It
has made the best possible use of the data.  But here it would have
been \emph{equally good}---and we would argue \emph{more useful}---to
produce a piece of shared \emph{code} that knows about these things
than an immutable data file or database that knows about these things:
The code would be readable (self-documenting), re-usable, and
modifiable.  Code \emph{passes on} knowledge, whereas a catalog
\emph{freezes} it.

In fact, a sufficiently persistent astronomer \emph{can} get the code
that was used to create the SDSS Catalogs, or at least parts of it.
But there is still a problem in principle: The SDSS Catalogs (in
common, as far as we know, with \emph{all} widely used astronomical
catalogs at the present day) have no \emph{justified probabilistic
description} or basis in the subject of \emph{statistical inference}.

Both Abell and the SDSS Collaboration used their catalogs to describe
a set of imaging data.  How do we judge whether or not a particular
description is a \emph{good} description?  If we have two reasonable
but different descriptions of the same imaging data, how can we choose
between them?  We need a quantitative measure of the \emph{quality} of
the description.  Fortunately, there is a standard solution to this
problem, which is justified in the context of Bayesian (or
frequentist) inference: The best description of the data is the one
that generates the highest-posterior-likelihood \emph{model} of the
data (or, if you like, of derived features, derived from the data by a
similar inferential process).

These issues are not theoretical: For example, in the SDSS Catalog,
there is a sophisticated and extremely well thought-out system that
performs ``deblending'' of overlapping sources.  This system performs
well.  However, it makes ``hard'' (binary) decisions about what
extended SDSS sources are groups of overlapping sources, and it
separates those sources into components according to those hard
decisions.  An investigator who wants to question those
decisions---decisions made by the code---has no way to perform any
kind of hypothesis test between a catalog with a certain source
deblended and another catalog that is very similar but has that same
source \emph{not} deblended.  This situation comes up frequently when
the SDSS Catalog is compared to data from other telescopes, where
sometimes the comparison data decisively contradict the SDSS Catalog
on a point of deblending, and the investigator is squeezed between
either ignoring the new data (that is, sticking with the SDSS
deblending decision despite the new data) or else re-analyzing the
SDSS imaging data from scratch.

Imagine now that along with the SDSS Catalog, the Collaboration had
released a piece of code that converts any contiguous chunk of the
catalog into synthetic images or model images of that patch of sky,
and---if requested---returns a likelihood or posterior probability of
those model images given the imaging data and the noise model.  This
joint release of catalog and code would provide users with a range of
new capabilities that go far beyond what is possible with the current
Catalog:
\begin{itemize}
\item Error analyses could be performed by adjusting the catalog
entries and re-evaluating the likelihood.  This ``non-parametric''
approach to error analysis allows non-trivial (for example,
non-Gaussian and non-Poisson) error estimates to be computed and
propagated easily.
\item It would be possible to measure \emph{any} element of the full
``billion by billion'' covariance matrix of catalog parameters by
adjusting the two catalog entries at the same time and re-evaluating
the likelihood.
\item Catalogs from different imaging data sets (with different
resolutions and at different wavelengths) could be ``matched'' or
compared or fit jointly by finding the catalog parameters that
optimize the product of likelihoods of the two data sets.  (This
presumes that the ``language'' in which catalogs are communicated is
sufficiently flexible that, for example, a single catalog could
be submitted to the HST, SDSS, and GALEX ``synthetic image'' systems
without error.)  If that became possible, then the entire field of
``catalog matching'' and the complexity found
therein~\cite{matching}---including issues like the deblending issue
mentioned above---could disappear.
\item Catalog entries could be made self-documenting, in the sense
that it would be easy for investigators to see the effect in the
synthetic images of making variations in the catalog parameters.
\item Idiosyncratic photometric parameters or other metrics not
specifically measured by the catalog makers (and there are many such
requests made of the SDSS Catalog makers, even after the Catalog
contains dozens of measurements per object per band) could be
computed---approximately---by applying the operation to the synthetic
images; and the same code could be used to compute the quantity
exactly by retrieving the imaging data.
\end{itemize}

In addition, a ``catalog plus code'' approach gives catalog builders a
way of expressing the ambiguities in the catalog.  Rather than making
a hard decision about each catalog entry, the catalog builders can
make ``soft'' decisions by trying each reasonable alternative of each
decision and producing a \emph{set} of possible catalog entries that
explain a particular image region.  These samples can be
weighted in a principled way: By using their likelihoods, the weight
of a potential catalog entry corresponds to its ability to explain the
observed image.  In this setting, the catalog builders no longer have
to set arbitrary thresholds for hard decisions, but instead must set
prior probabilities to bias soft decisions.  This clarifies the role of
priors in the construction of the catalog, and permits
other investigators to inject side information.

Finally---and we will return to this below---there is a direct
relationship between (lossless) compression of data and Bayesian
inference such that---if you choose the right encoding---the most
posterior-probable model of a data set also provides the best possible
lossless compression of those data for transmission over a
channel~\cite{inference}.  In this context, the best possible lossless
compression of, for example, the SDSS data, is---in the optimistic
future in which the ``catalog'' is a model of the data---the catalog,
the code that converts that catalog into synthetic images, and a map
of the residuals.  This picture gives new meaning to the idea of a
catalog being the result of ``data reduction''.

Of course there are many practical issues, since (as with any
catalog-building), the code needs to know quite a bit about the
imaging and sensitivity properties of the data set, and have a
realistic noise model for computing likelihoods.  In what follows, we
are going to make a general proposal for such a system, deduce some of
its potential capabilities, and give some sense of what baby steps we
are taking towards implementation.

Don't get us wrong: we love the SDSS Catalogs; one of us (DWH) was
involved in a minor way in their construction and
vetting~\cite{hoggsdss}.  Going forward, we are hoping we can do even
better.

\section{Calibration and standardization}

We seek a catalog---which, as we have argued, is an image model---that
is a good description not just of one image or one set of images, but of
\emph{every astronomical image ever taken}.  We need these images to
be vetted and calibrated in a consistent way, and we need the vetting
and calibration meta-data to be computed and stored in a consistent
way for every image.  This motivates fully automated vetting and
calibration.

We have built a successful system in this
domain---\textsl{Astrometry.net}~\cite{astrometrynet}---a reliable and
robust system that takes as input an astronomical image, and returns
as output the pointing, orientation, and field-of-view of that image
(the astrometric calibration).  The system requires no first guess,
and works with the information in the image pixels alone.  The success
rate is $>99.9$~percent for contemporary near-ultraviolet
(\textsl{Galaxy Evolution Explorer}~\cite{galex}) and visual
(SDSS~\cite{sdss}) imaging survey data, with \emph{no false
positives}.  We are using this system to generate consistent and
standards-compliant meta-data for all public digital and digitized
astronomical imaging from plate repositories, individual scientific
investigators, and amateurs.

Our basic approach involves two components.  The first is an indexing
system for asterisms, which can examine a query image and very quickly
generate candidate explanations for that image, where each explanation
consists of a proposed location, orientation and field-of-view on the
celestial sphere.  The second component is a verification criterion
that judges any proposed explanation.  Calibration proceeds by
extracting the two-dimensional positions of stars in a query image,
typically yielding a few hundred stars localized to pixel accuracy or
better.  Then, using subsets of stars (asterisms, usually of four
stars, but the system is general), the indexing system generates
\emph{hypotheses}---proposed alignments of the query image to the sky.
We assign a score to each hypothesis by measuring its ability to
predict the locations of other stars in the image (taking into account
that we cannot expect the image and catalog stars to overlap exactly).
This ``score'' is a well-posed probability, under some justifiable
assumptions; it permits us to reasonably estimate the chance of the
hypothesis being a false match; we output a result only when this
chance is vanishingly small (of order $10^{-9}$).  We continue
generating and testing asterism-inspired hypotheses until we find one
with sufficient confidence and output it as a confident match and
therefore calibration.  In some cases, we never find a hypothesis with
sufficient confidence (or we give up searching before we find one),
but our thresholds are set such that we essentially never produce a
false positive match.

The astrometric calibration locates the image on the sky, and also
identifies sources in the image with known sources in established
catalogs.  This permits other kinds of calibration.  We have shown,
for example, that comparison of the precise positions of the stars in
the image with their positions and proper motions tabulated in current
catalogs determines the \emph{date} at which the image is taken, to an
accuracy of years in typical data~\cite{blinddate}.  We are currently
working on other kinds of calibration:
\begin{itemize}
\item Analysis of sources known in previous catalogs to be stars
provides an estimate of the point-spread function and its variation
over the field.  Right now there is no widely agreed-upon method for
storing or communicating point-spread function meta-data, but
certainly this is essential.
\item Comparison of the brightness ordering of sources in the image
with brightness ordering in other multi-band catalogs provides a
reliable estimate of the broad-band wavelength bandpass.
\item Comparison with tabulated magnitudes or flux densities provides
an estimate of the sensitivity or zero-point.
\item Statistics of the dark and bright parts of the image can be used
to infer aspects of the noise model and saturation or nonlinearity
effects.  Some of these can be pathological, even for science-grade
data.  Since right now these kinds of problems are usually handled
with expert systems built on observer ``folklore'', this is the area
in which we have the least confidence in our ability to work
automatically.
\end{itemize}

We seek to remove astronomers from the calibration step.  Of course
one problem with the limited approaches we are taking with
\textsl{Astrometry.net} is that the system is taking the images
\emph{independently} and not using the \emph{joint} information about
calibration from many images that can far exceed the information in
any individual image.  Think, for example, of all the images from the
SDSS Telescope; we understand the world coordinate systems and its
variations and the point-spread function and its variations much
better by considering all of the images \emph{simultaneously} than we
do by considering them one at a time.

The quality of system we need for the ``theory of everything'' is such
that it must be able to \emph{discover} these groups of similar and
related data, and discover that there are synergies to the joint
analysis.  (There is an alternative, which is to ``hard-code'' these
synergies, but we are imagining an optimistic future in which this is
not possible given the eventual scale.)

Another problem with our current approaches is that we don't have good
strategies for calibrating truly raw data streams; that is, for
inferring the pixel-by-pixel relative sensitivities, for inferring
additive components like bias and zero-exposure data (think hot pixels
and charge sources).  These steps \emph{require} that we successfully
group input data by source and use it jointly.  Similarly, we don't
have good strategies for inferring non-trivial noise models or
saturation and nonlinearity problems.  There are many important
problems here, all of which must be solved if we are going to model
the entirety of the available data.

Finally, we have a problem of \emph{reliable data}.  Our approach
assumes that all data are equally unreliable, and must be calibrated
from scratch by our robots.  For some data sets this may be
unnecessary or even counter-productive, where instrument or survey
calibration teams have done a wonderful job.  How do we detect these
situations automatically---and inject correct priors---or do we really
have to calibrate everything from scratch?  Will automatic calibration
\emph{really} be better than calibration by the best craftspeople?
What we have going for us is that CCDs (and equivalent detectors) and
telescopes have a limited range of properties and faults.  What we
have going against us is that there are some beautifully understood
telescopes and cameras, calibrated as well as their entire data
outputs allow.

We have emphasized \emph{scale} here, but we also need automated
calibration to enforce \emph{standards}: The procedures by which
calibration is performed, and the language in which the calibration
meta-data are communicated, must be standardized across all the data
if joint simultaneous modeling is going to be successful and reliable.
This might argue for going with automated calibration \emph{even} in
situations where it costs us some precision.  That's a quantitative
question.

\section{Automated discovery}

In our world, a catalog is a probabilistic model of the data; indeed,
it \emph{generates} a synthetic version of every image upon which it
is based.  In our optimistic future, every new piece of imaging data
can be \emph{explained by} this catalog/model and is also used to
\emph{improve} this catalog/model.

The synthetic-image or generative aspect of this model relies
completely on reliable and standardized calibration meta-data for each
image.  Hence the relevance of our automated calibration efforts.
These meta-data, at a minimum, must include the astrometric
calibration (world coordinate system), the point-spread function as a
function of position, the bandpass and sensitivity (zeropoint) of the
image (also possibly as a function of position), and whatever is
necessary about flatfielding, saturation, defects, and other data
complexities.  Each incoming image can be synthesized by the catalog,
given these calibration meta-data.  Preferably, image synthesis
happens as close to the ``raw data'' (original instrument read-out) as
possible, but there is a well-posed version of this problem at the
``flat-fielded'' level also, provided that the calibration was found
by a justifiable inferential process.  Of course, in the long run, the
model comparison with the incoming image and the calibration of the
incoming image ought to be performed \emph{simultaneously}, of course.

The key idea is that the \emph{residuals} of the incoming image
against the catalog's synthetic image---the image pixels minus the
model image pixels---contain information that \emph{improves} the
model.  These improvements can take several form:
\begin{itemize}
\item The residuals could suggest making small adjustments to the
positions, fluxes, morphologies, or time-dependence of known sources.
Small adjustments can be made to existing model parameters so that the
overall posterior probability of the entire data set (prior data plus
incoming data) is improved.
\item A source that was previously modeled as not moving or not
varying or of some particular type could produce residuals in the
incoming image that are dramatically reduced if the source is
permitted to move or vary or be of another type.  New parameters can
be spawned in the model that provide the freedom to make these changes
and then be set to values that optimize the overall posterior
probability.
\item A concentrated collection of positive residuals (in image minus
model) could be found to be consistent with the morphology of a
point-source (or typical extended source, but inconsistent with being
a cosmic ray or a detector defect) in the incoming image.  A new
source can be added to the model to explain these residuals, and its
parameters can be set to optimize the overall posterior probability,
perhaps also simultaneously adjusting some of the parameters of nearby
or overlapping sources.
\item The residuals in the incoming image, when considered \emph{in
concert} with the residuals in all of the other overlapping images,
might make it such that the posterior probability is improved by
adding a new \emph{faint} source.  This would be a source that is too
faint to detect in any individual image, but appears when one
considers the joint information in all of the overlapping images.
This could be a well-posed version of what astronomers do when they
``co-add'' their data (co-addition is rarely the right thing to do
with real data, which don't tend to have purely Poisson errors or be
uncontaminated by cosmic rays and other nonlinear defects).
\end{itemize}

Notice that all four of these scenarios are scenarios of ``automated
discovery''.  The first is least interesting, because the
``discovery'' is simply that the model needs a tweak.  However, the
following three count as true discoveries, in that the inclusion of
the incoming image has made a \emph{qualitative change} to the
catalog, and therefore a qualitative change to our description of the
data.  The theory of everything is a \emph{framework for automated
discovery}.

Our description makes many mentions of the ``posterior probability''.
This is the probability that is obtained by multiplying, in a Bayesian
scheme, the prior probability of the model by the likelihood of the
data given the model (and normalizing sensibly).  Here the priors
matter, deeply: After all, how do you compare one model that does a
good job, and another that does a better job, but requires more stars,
or more of the stars to move?  The frequentist answer is to use
$\chi^2$ (the total of the squared residuals in units of the
uncertainties) divided by the number of degrees of freedom (the number
of data points less the number of parameters).  This is not insane,
but it is only truly justified in the case of a linear model and
Gaussian errors.  We don't have \emph{either} of these in general,
certainly not here.

We advocate a \emph{communication} prior: We prefer the model that,
when put into a suitably-encoded message along with the residuals of
the image against the model, makes for the shortest total message.
The posterior probability is, in this case, essentially the total
number of bits required to communicate the model parameters \emph{and}
the residuals.  A better model is better either because it reduces the
dynamic range of the residuals (so the residuals require fewer bits)
or because it reduces the number of parameters (so the model requires
fewer bits).  In the case of a linear problem with Gaussian errors,
this reduces to the model that minimizes $\chi^2$ per degree of
freedom, more-or-less.  We prefer this kind of prior for the simple
reasons that it has \emph{some} justification in terms of \emph{data
reduction}, it reduces to the frequentist answer in the trivial case,
and it is easy to apply in real situations.  It implements what is
essentially the ``minimum message length
criterion''~\cite{messagelength} for Bayesian model selection.

We have taken a step towards this kind of automated discovery with a
small project in the SDSS Southern Stripe~\cite{sdssss}, where the
SDSS Telescope has imaged $300~\mathrm{deg}^2$ repeatedly---30 to 50
times---in five bands over five years, and produced a calibrated time
series of images.  We performed image modeling of extremely red
sources that are detected in the combined data but too faint to be
detected in any individual epoch~\cite{faintmotion}.  We find, as
expected, that we can reliably determine proper motions of these faint
sources despite the fact that we can't centroid them at any epoch.
The moving sources we find are consistent with being extremely
low-mass sub-stellar objects, discovered ``automatically'' in the
sense we have used it here.

The larger problem---the theory of everything---is an immense problem
in data management, calibration, and optimization.  However, the model
and the data are both extremely ``sparse'': Widely separated parts of
the sky have almost completely independent parameters, and images from
different cameras on different telescopes share very little in their
calibration.  This means that the model can, in principle, be updated
in a local or atomic way.  We are confident that---without substantial
new technology and in a timescale of years (not decades)---the
entirety of preserved, digitized or digital astronomical imaging
\emph{can} be assembled and calibrated, and we are confident that an
enormously parameterized model of the type we envision \emph{can} be
adjusted to optimize a posterior probability that is sensitive to
every pixel in that imaging.  This, truly, would be an astronomer's
``theory of everything''.

\begin{theacknowledgments}
  It is a pleasure to thank Jo Bovy, Rob Fergus, Hans-Walter Rix, and
  Sam Roweis for valuable discussions and fruitful collaboration.
  This project was partially supported by the US National Science
  Foundation (grant AST-0428465) and the US National Aeronautics and
  Space Administration (grants NAG5-11669 and 07-ADP07-0099).  For
  part of this project, DWH was a research fellow of the Alexander von
  Humboldt Foundation in residence at the Max-Planck-Institut f\"ur
  Astronomie in Heidelberg.
\end{theacknowledgments}

\end{document}